\documentclass[aps,twocolumn,a4paper,pra,nofootinbib]{revtex4}
\usepackage{epsfig}
\usepackage{amssymb}
\usepackage{latexsym}
\usepackage{amsfonts}

\def\beq{\begin{eqnarray}}
\def\eeq{\end{eqnarray}}
\def\kB{k_{\rm B}}
\def\lP{\ell_{\rm P}}
\def\lC{\ell_{\rm C}}
\def\tP{t_{\rm P}}
\def\mP{m_{\rm P}}
\def\Hz{{\rm Hz}}
\def\K{{\rm K}}
\def\gw{{\rm gr}}
\def\d{{\rm d}}
\def\s{{\rm s}}
\def\m{{\rm m}}
\def\g{{\rm g}}
\def\at{{\rm at}}
\def\phot{{\rm phot}}

\def\Sagnac{{\rm Sagnac}}
\def\LT{{\rm LT}}
\def\Moon{{\rm Moon}}

\begin{document}

\title{HYPER and gravitational decoherence}
\author{Serge Reynaud}
\homepage{www.spectro.jussieu.fr/Vacuum}
\author{Brahim Lamine}
\author{Astrid Lambrecht}
\affiliation{Laboratoire Kastler Brossel, case 74, 
Campus Jussieu, 75252 Paris, France}
\thanks{Laboratoire du CNRS, de l'ENS 
et de l'Universit{\'e} Pierre et Marie Curie} 
\author{Paulo Maia Neto}
\affiliation{Instituto de F\'{\i}sica, UFRJ, 
CP 68528, 21945-970 Rio de Janeiro, Brazil}
\author{Marc-Thierry Jaekel}
\affiliation{Laboratoire de Physique Th\'{e}orique, 
24 rue Lhomond, F75231 Paris, France}
\thanks{Laboratoire du CNRS, de l'ENS, de l'UPMC 
et de l'Universit{\'e} Paris Sud} 

\begin{abstract}
We study the decoherence process associated with 
the scattering of stochastic backgrounds of gravitational waves.
We show that it has a negligible influence on
HYPER-like atomic interferometers although it 
may dominate decoherence of macroscopic motions, such as the 
planetary motion of the Moon around the Earth. 
\end{abstract}

\maketitle


\section{Introduction}	

Decoherence is a general phenomenon which occurs 
for any physical system coupled to any kind of environment. 
It plays an important role in the transition between microscopic 
and macroscopic physics by washing out quantum coherences on a time 
scale which becomes extremely short for systems with a large degree 
of classicality or, in other words, by suppressing 
superpositions of different quantum states when the latter
have sufficiently different classical properties 
\cite{Zeh70,Dekker77,Zurek81,Caldeira83,Joos85}. 

For large macroscopic masses, say the Moon orbiting around the Earth,
decoherence is in fact so efficient that the classical description 
of the motion is sufficient. Precisely, the decoherence time scale 
is so short that the observation of any quantum coherence is impossible. 
For microscopic masses in contrast, decoherence is expected to be so 
inefficient that we are left with the ordinary quantum description
of the system. If we consider for example electrons orbiting inside 
atoms, the decoherence time scale is so long that decoherence can be 
forgotten. 

Decoherence has been observed in a few experiments only and this can be 
understood from the simple arguments sketched in the previous paragraphs.
Decoherence can only be seen by dealing with `mesoscopic' systems for 
which the decoherence time is neither too long nor too short. 
The micro/macro transition has then to be assessed by following the 
variation of this decoherence time with some parameter measuring 
the degree of classicality of the system. 
These experimental challenges have been met with 
microwave photons stored in a high-Q cavity \cite{Brune96,Maitre97} 
or trapped ions \cite{Myatt00}.
In such model systems where the fluctuations are particularly
well mastered, the quantum/classical transition has been shown 
to fit the predictions of decoherence theory \cite{Raimond01}.

It has been suggested that matter-wave interferometers could reveal 
the existence of intrinsic spacetime fluctuations through decoherence 
processes \cite{Percival97,PercivalS97,Power00,Amelino99,Amelino00}.
The effect has not been seen in existing matter-wave interferometers
\cite{Peters99,Gustavson00} but more sensitive instruments are
now being developed, like the atomic interferometer HYPER designed to
measure the Lense-Thirring effect in a space-borne experiment 
\cite{Hyper00,Hyper02} and it is important to obtain quantitative 
estimates of the effect of decoherence associated with spacetime
fluctuations for such instruments.

The perturbations of interest in the study of an atomic interferometer
correspond to frequencies much smaller than Planck frequency. At such
frequencies, general relativity is an
accurate effective description of gravitation, although it is
certainly not the final word \cite{Will90,Damour94}. 
It follows that the intrinsic spacetime fluctuations which constitute 
our gravitational environment are essentially the free solutions of 
general relativity, that is also the gravitational waves predicted by 
the linearized form of the theory \cite{Weinberg65,Grishchuk77,Zeldovich86}. 
This linearized form is widely used for studying propagation of gravitational waves 
and their interaction with the presently developed interferometric detectors
\cite{Schutz99,Maggiore00,Ungarelli00,Grishchuk01}.

In the present paper, we will study the decoherence of atomic interferometers due 
to their interaction with the stochastic background of gravitational waves emitted 
by astrophysical or cosmological processes \cite{Lamine02}. 
We will show that this scattering does not lead to an appreciable decoherence 
for the atomic interferometers presently studied, HYPER being chosen as the
typical example. Incidentally, this ensures that HYPER will not have its interference 
fringes destroyed by decoherence and will therefore be able to measure Lense-Thirring effect.
We will contrast this answer with recent results showing that 
the gravitational decoherence is the dominant decoherence mechanism, and
an extremely efficient one, for macroscopic motions such as the motion of the 
Moon around the Earth \cite{Reynaud01}.  

This contrast is directly connected to a dimensional argument~: 
since gravity is coupled to energy, the associated decoherence effects are certainly 
more efficient for macroscopic masses than for microscopic ones. In particular, the 
mass of the Moon is larger than Planck mass by orders of magnitude whereas 
the atomic probes used in HYPER have their mass much smaller 
than Planck mass. Of course, this scaling argument is not by itself sufficient 
to answer quantitative questions about the decoherence rates. We will give 
below precise estimations of the gravitational decoherence effect which
depend not only on the mass of the atoms, but also on their velocity,
on the geometry of the interferometer and on the noise spectrum
characterizing the gravitational fluctuations in the relevant frequency
range.

\section{Gravitational waves and atomic interferometers}	

Gravitational waves, as well as other gravitational perturbations such as 
the Lense-Thirring frame-dragging effect, dephase the matter waves in the 
two arms of the interferometer and thus affect the interference fringes. 
As a consequence of their stochastic character, the dephasings result in a loss
of contrast of the interference fringes when averaged over the integration
time of the measurement. We will use these ideas below to evaluate the effect 
of decoherence and write it in terms of properties of the interferometer 
on one hand and of the gravitational fluctuations on the other hand.
In the present paper, we will focus the attention on simple qualitative 
descriptions of the effect of gravitational waves on an atomic interferometer 
such as HYPER. The reader interested in more detailed discussions is referred to 
other contributions printed in the present volume \cite{Hyper02} or to
reviews on atomic interferometry \cite{Borde00,Borde01,Peters01}. 
More details on the interaction of gravitational waves with HYPER-like
interferometers may also be found in \cite{Lamine02}.

\begin{figure}[t]
\epsfig{file=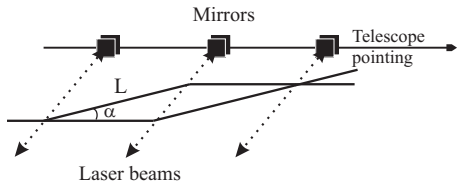,width=6cm}
\caption{Schematic representation of an HYPER-like atomic interferometer.
\label{fig:Hyper}}
\end{figure}

We consider the simplest description of HYPER as a matter-wave interferometer
with a rhombic symmetry (see Fig.\ref{fig:Hyper}). In a first step, we discuss
only the dephasings of matter waves and disregard the laser beams used as
beam splitters or mirrors in the interferometers. We will come back to the
dephasings of these laser beams later on.

We also restrict the discussion on this interferometer used as a gyrometer
and ignore acceleration effects. The archetype of a measurement performed in this
manner is the measurement of the rotation of the interferometer versus inertial
frames through the observation of Sagnac effect. When non relativistic atoms
are used, the Sagnac dephasing $\Phi_\Sagnac$ is found to be proportional 
to the mass $m_\at$ of the atoms, the area $A$ of the interferometer
and the rotation frequency $\Omega_\Sagnac$ 
\beq
&&\Phi_\Sagnac = \frac{ 2 m_\at A} {\hbar} \Omega_\Sagnac  
\eeq
The area is given by the length $L$ of the rhomb side and the
aperture angle $\alpha$ (see Fig.\ref{fig:Hyper}) 
\beq
&&A = L^2 \sin\alpha = v_\at^2 \tau_\at^2 \sin\alpha
\eeq
Alternatively, $L$ may be substituted by the product $v_\at \tau_\at$ where 
$v_\at$ is the atomic velocity and $\tau_\at$ the time of flight on one rhomb side.

The Sagnac effect measures the rotation frequency of the atomic interferometer
with respect to the inertial frame as it is defined at its location.
In general relativity, this local inertial frame differs from the celestial
frame determined by the `fixed stars' as a consequence of the dragging 
of inertial frames by the rotation of nearby bodies. 
This gravitomagnetic Lense-Thirring effect can be observed by comparing the
local inertial measurement performed by the atoms to the indication of the star 
tracker pointing to a star. 
It can be described by a dephasing $\Phi_\LT$ 
written analogously to the Sagnac dephasing $\Phi_\Sagnac$ 
\beq
&&\Phi_\LT  = \frac{2m_\at A} {\hbar} \Omega_\LT  
\eeq
The rotation frequency $\Omega_\LT$ measures the frame dragging induced,
for the HYPER project, by the rotation of the Earth. It is given by general
relativity and depends on the position of the satellite on its orbit but
it does not vary with time at a fixed spatial location.

We now use these reminders to describe in simple words the coupling of the 
interferometer to stochastic gravitational waves. These waves are
consequences in our local environment of the motion of masses in the Galaxy
or, more generally, in the Universe. In fact, they are the radiation fields,
freely propagating far from their sources, originating from the gravitomagnetic
fields present in the vicinity of the sources. Their effect on
the gyrometer may be written analogously to the Lense-Thirring dephasing
\beq
&&\delta \Phi_\gw = \frac{2m_\at A} {\hbar} \delta \Omega_\gw 
\label{Phigw}
\eeq
This analogy must not be pushed too far~: the Lense-Thirring field is a 
quasistatic near field while the gravitational waves are radiated far fields.
The dephasing $\delta \Phi_\gw$ and frequency $\delta \Omega_\gw$ 
are stochastic variables representing a time-dependent dragging of the local
inertial frame. 

To be more precise, we may write the dephasing $\delta \Phi_\gw$ due to the 
effect of gravitational waves as they are registered by the atoms along the
two arms of the interferometer. We disregard most perturbing effects which 
play a role in the real interferometer \cite{Hyper02}. 
We treat only the dominant contribution to the dephasing $\delta\Phi _\gw$, 
that is the difference between the phases accumulated on the two arms 1 and 2 
by slow atoms because of the geodesic perturbation \cite{Linet76}
\beq
\delta\Phi _\gw (t) = \frac{m_\at}{2\hbar} &&\left[ 
\int_{1}  
\ h_{\rm ij} (t^\prime) v^{\rm i} (t^\prime) v^{\rm j} (t^\prime) \d t^\prime 
\right. \nonumber \\
&&-\left. \int_{2}  
\ h_{\rm ij} (t^\prime) v^{\rm i} (t^\prime) v^{\rm j} (t^\prime) \d t^\prime 
\right] 
\eeq
Here the metric components are evaluated in a specific transverse traceless gauge
(see below).
The dephasing can also be written in an explicitly gauge invariant manner by
taking into account all the components of the interferometer, in particular
the mirrors. For the purpose of the present paper, we do not need to enter into 
these subtle descriptions. 

\begin{figure}[t]
\epsfig{file=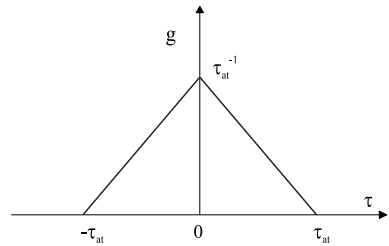,width=5cm}
\caption{Filtering function asociated with the atomic gyrometer 
of Fig.\ref{fig:Hyper}~: it describes the averaging of the 
rotation frequency associated with the finite
time of flight of atoms. 
\label{fig:Gdetau}}
\end{figure}

Using the symmetry of the rhomb, it is easy to rewrite this expression as in 
equation (\ref{Phigw}) with the equivalent rotation frequency $\delta\Omega _\gw$ 
obtained from the derivative of the metric component $h_{\rm 12}$
\beq
&&\delta\Omega _\gw (t) = -\frac 12 \frac{\d \overline{h_{\rm 12}}}{\d t} 
\eeq
The directions 1 and 2 correspond to the spatial plane defined
by the interferometer. The averaged quantity $\overline{h_{\rm 12}}$ is obtained 
from the metric component $h_{\rm 12}$ through a convolution 
\beq
&&\overline{h_{\rm 12}}(t) = \int \ h_{\rm 12} \left(t - \tau \right) 
g\left(\tau\right) \d \tau
\eeq
The linear filtering function $g$ is represented on Fig.\ref{fig:Gdetau},
with a triangular shape which reflects the distribution of the time of exposition 
of atoms to gravitational waves inside the rhombic shape of the interferometer.
It differs from 0 for values of $\tau$ having a modulus smaller than $\tau_\at$
and has an integral normalized to unity.
Its Fourier transform describes the linear filtering in frequency space 
\beq
&&\tilde{g} \left[ \omega \right] = 
\left( \frac { \sin \frac{\omega \tau_\at}{2} } 
{ \frac{\omega \tau_\at}{2} } \right) ^2 
\eeq
The square of this function is the apparatus function discussed in \cite{Lamine02}.

\section{Gravitational wave backgrounds}	

We now explain how we describe the fundamental fluctuations of space-time 
and their effect on the motion of matter.
The basic idea is that the frequency range of interest lies far below Planck 
frequency, for all systems of current experimental interest.
At these frequencies, general relativity is an accurate description of 
gravitational phenomena, and this statement is essentially independent of the 
modifications of the theory which will have to take place in a complete 
theory of quantum gravity \cite{Jaekel95}. 

It follows that the intrinsic spacetime fluctuations which constitute
our gravitational environment are simply the gravitational waves 
predicted by the linearized version of Einstein theory of gravity
and which are thoroughly studied in relation with the ongoing experimental 
development of gravitational wave detectors. 
In this context, the stochastic backgrounds of gravitational waves emitted by 
astrophysical or cosmic processes are of particular interest.
These backgrounds might be treated in a gauge-invariant manner by introducing 
the correlation functions describing the fluctuations of curvature. 
For the sake of simplicity, we will present here evaluations in a specific gauge.
We choose the transverse traceless (TT) gauge which is tangent to the proper frame
of the atomic interferometer at the time $t$ of measurement.
In this gauge, often used in the studies of gravitational wave detectors,
metric components vanish as soon as they involve a temporal index 
\beq
&&h_{\rm 00} = h_{\rm 0i} = 0 
\eeq
(i,j=1,2,3 stand for the spatial indices whereas 0 will represent the temporal 
index); the spatial components $h_{\rm ij}$ of the metric tensor 
are directly connected to the Riemann curvature or, equivalently in free space, 
to the Weyl curvature
\beq
&& \frac{\d^2 h_{12}}{\d t^2} = -2 R_{1020} = -2 W_{1020} 
\eeq
so that the trace of the spatial components vanish
\beq
&&h_{\rm ii} = 0 
\eeq

Then the gravitational waves are conveniently described through a
mode decomposition 
\beq
&&h_{\rm ij}\left( x\right) = \int \frac{{\rm d} ^4 k}{\left( 2\pi \right) ^4}
\ h_{\rm ij}\left[ k\right] e ^{ -ik_\mu x^\mu }  
\eeq
Each Fourier component is a sum over the two circular polarizations $h^+$ 
and $h^-$ 
\beq
&& h_{\rm ij} \left[ k\right] = \Sigma _\pm 
\left( \frac {\varepsilon _{\rm i}^{\pm} \varepsilon _{\rm j}^{\pm}} {\sqrt{2}}
\right) ^{*} h^{\pm}\left[ k\right] 
\eeq
The gravitational polarization tensors are obtained as products of the polarization 
vectors $\varepsilon _{\rm i}^{\pm}$ well-known from electromagnetic theory.
The gravitational waves correspond to wavevectors $k$ lying on the light cone
($k^2=0$), they are transverse with respect to this wavevector
($k^{\rm i} \varepsilon _{\rm i}^{\pm} =0$) and the metric perturbation
has a null trace.

We consider for simplicity the case of stationary, unpolarized and isotropic 
backgounds. Then, a given metric component, say $h \equiv h_{12}$,
evaluated at the center of the interferometer as a function of time $t$,
is a stochastic variable entirely characterized by a noise spectrum $S_{h}$ 
\beq
&&\left\langle h \left( t \right) h \left( 0 \right) \right\rangle 
= \int \frac{{\rm d} \omega}{2\pi}
\ S_{h}\left[ \omega \right] e ^{ -i\omega t }  
\eeq
$S_{h}$ is the spectral density of strain fluctuations considered in most
papers on gravitational wave detectors (see for example \cite{Maggiore00}). 
It has the dimension of an inverse frequency.
It can be written in terms of the mean number $n _\gw $ of gravitons per mode 
\beq
&& S_{h} = \frac{16 G}{5 c^5} \ \hbar \omega n _{\gw} 
\eeq
or, equivalently, of a noise temperature $T _{\gw}$ 
\beq
&&S_{h} = \frac{16 G}{5 c^5} \ \kB T _{\gw} 
\label{defTgw}
\eeq
with $\kB$ the Boltzmann constant and $G$ the Newton constant. 

\begin{figure}[t]
\epsfig{file=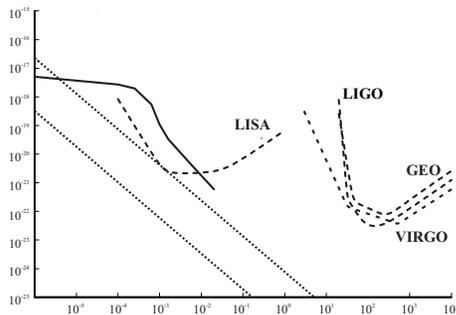,width=6cm}
\caption{Variation of the square root $\sqrt{S_h}$ of the spectral density of 
strain fluctuations (measured in Hz$^{-\frac 12}$) versus gravitational wave 
frequency $f=\frac{\omega}{2\pi}$ (measured in Hz)~: 
the dashed lines represent the sensitivity curves of detectors on ground 
(GEO, LIGO, VIRGO) and in space (LISA); the solid line describes the binary 
confusion background and the dotted lines represent potential cosmic backgrounds
with different parameters.  
\label{fig:Bckgrnd}}
\end{figure}

We have represented on Fig.\ref{fig:Bckgrnd} a part of the information 
available from the studies devoted to interferometric detectors of gravitational 
waves (see for example \cite{Schutz99}). 
The dashed lines represent the sensitivity curves for detectors~: 
the 3 curves on the right correspond to detectors presently built on ground 
with their optimal sensitivity in the 10Hz-10$^4$Hz;
the curve in the central part corresponds to the space project LISA with its optimal 
sensitivity in the 10$^{-4}$Hz-10$^{-1}$Hz. 
The solid line on the left part represents the `binary confusion background', that is
the estimated level for the background of gravitational waves emitted by unresolved 
binary systems in the galaxy and its vicinity.
This `binary confusion background' corresponds to a nearly flat function $S_{h}$,
that is also to a nearly thermal spectrum, in the $\mu\Hz$ to 10mHz frequency range 
\beq
10^{-6} \Hz < \frac \omega {2\pi} < 10^{-4}\Hz \ && \
S_{h} \sim 10^{-34} \Hz^{-1}
\eeq
With the conversion factors given above, this corresponds 
to an extremely large equivalent noise temperature 
\beq
&&T _{\gw} \simeq 10^{41}\ \K 
\eeq
It is worth stressing that $T _\gw $ is an effective noise 
temperature, that is an equivalent manner for representing 
the noise spectrum $S_{h}$, but certainly not a real temperature. 
The value obtained here for this temperature is much higher than the 
thermodynamical temperature associated with any known phenomenon.
It is even larger than Planck temperature ($\sim 10^{32}\ \K$), 
which emphasizes its unconventional character from the thermodynamical point of view.
In fact, the motion of matter is so weakly coupled to the gravitation that it remains 
always far from the thermodynamical equilibrium. 

The estimations discussed here correspond to the confusion background 
of gravitational waves emitted by binary systems in our Galaxy or its
vicinity. They may be treated as stochastic variables because of
the large number of unresolved and independent sources.
As a consequence of the central limit theorem, they may even be considered to 
obey a gaussian statistics, a property which will be used later on. 
Since they rely on the laws of physics and astrophysics as they are known in our 
local celestial environment, they may be considered as granted sources 
of gravitational waves. 
There also exist predictions for gravitational backgrounds associated 
with a variety of cosmic processes \cite{Maggiore00}. 
These predictions are represented by the dotted lines on Fig.\ref{fig:Bckgrnd}.
They depend on the parameters used in the cosmic models and have a more 
speculative character than local astrophysical predictions. 
The associated temperatures vary rapidly with frequency but they 
are usually thought to be dominated by the confusion binary background 
in the frequency range considered thereafter. 

\section{Gravitational decoherence of atomic interferometers}	

We come now to the evaluation of decoherence of the interferometer
due to the scattering of stochastic gravitational waves.
Here again, we choose to present a simple description. Decoherence 
will be understood as a loss of fringe contrast resulting from
the averaging of stochastic dephasings. Precisely, stochastic gravitational 
waves with frequencies higher than the inverse of the averaging time
will be identified with the unobserved degrees of freedom which
are usually traced over in decoherence theory 
(see \cite{Raimond01} and references therein).
The phase dispersion approach used in the present paper is known to be
equivalent (see for example \cite{Imry90}) to the other approaches to 
decoherence and it is obviously well adapted to the description of 
interferometers where the phase is the natural variable.

The evaluation of decoherence is presented in a more detailed manner
in \cite{Lamine02}. Here we merely consider the degradation of fringe contrast
obtained by averaging over stochastic dephasings. Since $\delta \Phi_\gw$ 
is a gaussian stochastic variable, the degraded fringe contrast is read as 
\beq
&&\left\langle \exp \left( i\delta \Phi_\gw \right) \right\rangle =
\exp \left( -\frac{ \Delta \Phi_\gw^2 } {2} \right) 
\eeq
where $\Delta \Phi_\gw^2$ is the variance of $\delta \Phi_\gw$ 
\beq
&&\Delta \Phi_{\gw}^2 = \left\langle \delta \Phi_\gw^2 \right\rangle 
\eeq
Using the expression of $\delta \Phi_\gw$ in terms of the averaged time derivative 
of $h_{\rm 12}$ we write the variance $\Delta \Phi_\gw^2$ as an integral over 
the noise spectrum $S_h$
\beq
\Delta \Phi_{\gw}^2 = 4 \mu_\at ^2 
\int \frac{\d\omega}{2\pi} S_h\left[\omega\right] 
\frac{\left( 1 - \cos\left(\omega\tau_\at\right)\right)^2}{\omega^2} &&
\eeq
We have introduced a parameter $\mu_\at$ which has the dimension of
a frequency and is essentially determined by the kinetic energy of
the atoms and the aperture angle of the interferometer 
\beq
&&\mu_\at = \frac{2m_\at A} {\hbar \tau_\at^2} 
= \frac{2m_\at v_\at^2 \sin \alpha} {\hbar} 
\eeq

Using this integral expression, we can calculate the variance $\Delta \Phi_\gw^2$ 
for an arbitrary noise spectrum $S_h$. For the purpose of the present paper,
we obtain interesting results by considering the special case where
the spectrum $S_h$ is approximately flat. This corresponds to the assumption
of a thermal spectrum which, as already discussed, is met by the binary
confusion background on a significant frequency range. 
With this kind of white noise assumption, the variance is found to be 
proportional to the constant value of the noise spectrum $S_h$. It is read as
\beq
&&\Delta \Phi_\gw^2 = \mu_\at ^2 \ S_h \ 2 \tau_\at 
\eeq
where $\tau_\at$ is the time of exposition of atoms to gravitational waves
and $\mu_\at$ the typical frequency scale already discussed.

After the substitution in this estimation of the numbers corresponding to 
HYPER \cite{Hyper00}, we deduce that the decoherence of the interferometer
due to the scattering of gravitational waves is completely negligible
\beq
\Delta \Phi^2_\gw \sim  10^{-20} &\ & \ll 1
\eeq
This number corresponds only to the direct effect of gravitational waves
on the matter waves involved in the atomic interferometer.
We have also to take into account the dephasings which are picked up
by the laser fields involved in the stimulated Raman processes used
for building up beam splitters and mirrors for matter waves. 

These fields register gravitational waves on their flights from the lasers
to the atoms. The calculation of the corresponding dephasings is discussed
in \cite{Lamine02} and is not repeated here. The result of this
calculation can be written under the same form as previously
\beq
&&\Delta \Phi_{\gw}^2 [\phot] \simeq \mu_\phot ^2 \ S_h \ 2 \tau_\phot 
\eeq
$\mu_\phot$ is the laser frequency, that is also the frequency scale
corresponding to the kinetic energy of one photon; 
$\tau_\phot$ is of the order of the time of flight of photons from the 
lasers to the atoms (see a more precise discussion in \cite{Lamine02}),
that is also the time of exposition of photons to gravitational waves.

Using the numbers corresponding to HYPER \cite{Hyper00}, it turns out that
the contribution to decoherence of optical dephasings largely dominates the
contribution of atomic dephasings
\beq
&&\Delta \Phi_{\gw}^2 [\phot] \gg \Delta \Phi_{\gw}^2 [\at] 
\eeq
while nevertheless remaining negligible
\beq
\Delta \Phi^2_\gw \sim  10^{-12}  &\ & \ll 1
\eeq
As already evoked in the Introduction, this is good news for HYPER~: 
intrinsic spacetime fluctuations do not have the ability to wash out
the interference fringes. Should we have found a variance
$\Delta \Phi^2_\gw$ of the order of unity, or greater, it would have been
difficult to vary this value while controlling the associated effects in HYPER.

The negligible effect of decoherence can be discussed in an interesting 
alternative manner. The phase noise level associated with gravitational waves
may be measured as an equivalent displacement $\delta q$ of the mirrors which 
reflect the lasers. The noise spectrum corresponding to this equivalent noise
is simply approximated as
\begin{eqnarray}
&&S_q [\omega] \sim S_h L_\phot^2 
\sim 10^{-34} \left( \mbox{m} / \sqrt{\mbox{Hz}}\right)^2 
\end{eqnarray}
We have again used the numbers of HYPER with the length $L_\phot$ of the
optical arms of the order of 1m. This corresponds to a noise level 
$\sqrt{S_q} \sim 10^{-17}\mbox{m} / \sqrt{\mbox{Hz}}$ which is far beyond 
the vibration noise level $\sqrt{S_q} \sim 10^{-12} \mbox{m} / \sqrt{\mbox{Hz}}$ 
which is the target of the HYPER instrument. This means that the
phase noise induced by the scattering of gravitational waves is
completely negligible with respect to the phase noise corresponding to 
mechanical vibrations of the mirrors. In the real instrument, decoherence
is expected to be induced by the latter instrumental fluctuations rather
than by the former fundamental fluctuations.

\section{Gravitational decoherence of planetary motions}	

This does not mean that the scattering of gravitational waves always has
a negligible contribution to decoherence. 
To make this point clear, we now consider the case of macroscopic motions,
say the planetary motion of the Moon around the Earth.
This case is often chosen in introductory discussions on decoherence as
the archetypical system for which decoherence is so efficient that 
quantum fluctuations can certainly not be observed.
In these discussions, decoherence is often attributed to collisions
of residual gaz, to radiation pressure of solar radiation or, even, to the 
scattering of electromagnetic fluctuations in the cosmic microwave background. 
In fact, as discussed here, the decoherence of planetary motions is 
dominated by the scattering of stochastic gravitational waves present in our 
galactic environment.

The Earth-Moon system can be thought of as a giant gyroscope and its motion
is thus sensitive to the stochastic dragging of the inertial frame already
discussed in the previous sections. 
For the sake of simplicity, we consider only the case of a circular planetary 
orbit in the plane $x_{\rm 1}x_{\rm 2}$. We introduce the reduced mass 
$m=\frac{m_{a}m_{b}}{m_{a}+m_{b}}$ and the total mass $M=m_{a}+m_{b}$,
which are defined in terms of the masses $m_{a}$ and $m_{b}$ of the two bodies.
The radius $\rho$, that is the constant distance between the two masses, 
and the orbital frequency $\Omega$ are related to the masses
by the third Kepler law 
\beq
&&\rho ^{3}\Omega ^{2} = GM
\eeq
We may also use as characteristic parameters the tangential velocity $v=\rho \Omega$ 
and the normal acceleration $a = \rho \Omega ^2 = \frac {v ^2} \rho$.

For planetary systems, the effect of gravitational waves is conveniently
described as a perturbation coupling the quadrupole momentum of the system 
to the Weyl curvature tensor $W_{\rm i0j0}$.
Using this description, it can be shown that the effect of gravitational waves 
is in fact a Brownian force acting along the mean circular motion.
As a consequence, the Moon undergoes a momentum diffusion 
with the variance $\Delta p^2$ of the transfered momentum varying linearly 
with the time of exposition $\tau$ to gravitational waves 
\beq
&&\Delta p^2 = 2 D _\gw \tau 
\eeq
The momentum diffusion coefficient $D_\gw$ is obtained as \cite{Reynaud01}
\beq
&&D_\gw = m \Gamma _\gw \kB T_\gw 
\eeq
$T_\gw$ is the effective noise temperature of the gravitational background,
evaluated at twice the orbital frequency;
$\Gamma _\gw$ is the damping rate associated with the emission of 
gravitational waves by the planetary system 
\beq
&&\Gamma _\gw =\frac{32Gma^2}{5c^5}  
\eeq
where $a$ is the normal acceleration on the circular orbit.

These formulas bring together the Einstein fluctuation-dissipation 
relation on Brownian motion \cite{Einstein05} and the Einstein quadrupole
formula for gravitational wave emission \cite{Einstein18}.
As it is well known, the gravitational damping is so small for the Moon 
that it has a negligible effect on its mean motion 
\beq
&&\Gamma _{\gw} \approx 10^{-34}\ \s^{-1} 
\eeq
It is only for strongly bound binary systems that gravitational damping 
has a noticeable effect \cite{Taylor92}. 
For the Moon, it is not only small but much smaller than the damping due
to the scattering of electromagnetic radiation pressure or to the
interaction between Earth and Moon tides which is the dominant contribution
to damping \cite{Bois96} 
\beq
&&\Gamma _{\gw} \ \ll \ \Gamma _{\rm em} \ < \ \Gamma _{\rm tides}
\eeq
However, we show now that the gravitational mechanism dominates the
decoherence process.

\begin{figure}[t]
\epsfig{file=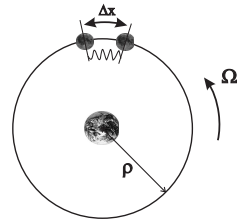,width=3cm}
\caption{Symbolic representation of the superposition of
two different motions of the Moon separated by a distance $\Delta x$ 
along the mean motion. In fact, any potential coherence between these 
two motions would be decohered in an extremely short time. 
The scattering of gravitational waves is the most efficient 
mechanism for this decoherence process.
\label{fig:Moon}}
\end{figure}

In order to evaluate decoherence, we consider two neighbouring 
motions on the circular orbit of the Moon around the Earth.
More precisely, we consider two motions characterized by the same 
spatial geometry but slightly different values of the epoch
- {\it i.e.} the time of passage at a given space point.
For simplicity, we measure the difference by the spatial distance $\Delta x$ 
between the two motions which is constant on a circular orbit
(see Fig.\ref{fig:Moon}).

As the gravitational wave perturbation depends on time,
these two motions undergo different diffusion processes. 
This differential effect has been evaluated in \cite{Reynaud01}
and we reproduce here the result of this evaluation.
Should we associate a quantum phase to a motion of the Moon, 
the two neighbouring motions would suffer a differential dephasing
characterized by an exponential $e ^{i \delta \Phi_\Moon}$.
We can then average this quantity over the stochastic effect of 
gravitational waves, still supposed to obey gaussian statistics. 
We obtain in this manner a decoherence factor 
\beq
&&\left\langle e ^{i \delta \Phi_\Moon} \right\rangle 
= \exp \left( -\frac {\Delta \Phi_\Moon^2} 2 \right) 
\eeq
which can be expressed in terms of the momentum diffusion coefficient,
of the time of exposition and of the distance between the two motions
\beq
&&\Delta \Phi_{\rm Moon}^2 = \frac{2 D _\gw \Delta x^2 \tau} {\hbar ^2} 
\eeq
This is just the result expected from general discussions on decoherence 
\cite{Zurek81} with decoherence efficiency increasing exponentially fast 
with $\tau$ and $\Delta x^2$.

For the sake of comparison with atomic interferometers, we rewrite this
formula by using (\ref{defTgw}) 
\beq
&&\Delta \Phi_{\rm Moon}^2 = \mu_{\rm Moon} ^2\ S_h \ 2\tau \\ 
\mu_{\rm Moon} &=& \frac{2mv^2}{\hbar} \sin\alpha \qquad 
\sin\alpha = \frac{\Delta x}{2\rho} \nonumber 
\eeq
$\mu_{\rm Moon}$ is a frequency determined by the kinetic energy of the Moon
and $\sin\alpha$ is the aperture angle of the equivalent interferometer.
With the numbers corresponding to the Moon, we find 
\beq
&&\frac{D_\gw}{\hbar^2} \approx 10^{75}\ \s^{-1}\m^{-2} 
\eeq
This corresponds to an extremely short decoherence time, even for
ultrasmall distances $\Delta x$. To fix ideas, the time lies in the 
$10\mu$s range for $\Delta x$ of the order of the Planck length.

In fact, the gravitational contribution to decoherence is found to be much larger 
than the contributions associated with tide interactions and electromagnetic
scattering
\beq
&&D_\gw \ \gg \ D_{\rm tides} \ > \ D_{\rm em} 
\eeq
The reversal of roles is due to the huge effective temperature
of the gravitational environment. To be more specific, the ratio 
$\frac{\Gamma_\gw}{\Gamma_{\rm tides}}$
of the damping constants associated with gravitational waves and
tides is a very small number of the order of $10^{-16}$.
But, at the same time the ratio $\frac{T_\gw}{T_{\rm tides}}$
is an extremely large number of the order of $10^{38}$.
It follows that the ratio $\frac{D_\gw}{D_{\rm tides}}$
is itself very large so that the gravitational contribution to
decoherence is found to dominate the other contributions.

This entails that the ultimate fluctuations of the motion of Moon, and the
associated decoherence mechanisms, are determined by the classical gravitation 
theory which also explains its mean motion. 
In other words, the environment to be considered when dealing with
macroscopic motions consists in the gravitational waves of the confusion
binary background.
This background is naturally defined in the reference frame of the galaxy
if it is dominated by galactic contributions or in a reference frame
built on a larger region of the universe if extragalactic contributions
have to be taken into account.

\section{Gravitational decoherence and Planck scales}	

The results obtained for gravitationally induced decoherence are 
reminiscent of the qualitative discussions of the Introduction. 
For macroscopic bodies, such as the Moon orbiting around the Earth,
decoherence is extremely efficient with the consequence that potential quantum 
coherences between different positions can never be observed. 
For microscopic probes, such as the atoms or photons involved in HYPER,
decoherence is so inefficient that it can be ignored with the consequence
that ordinary quantum mechanics can be used.

We remark that the Planck mass, that is the mass scale which can be 
built up on the constants $\hbar$, $c$ and $G$, lies on the borderland 
between microscopic and macroscopic masses
\beq
&& \mP = \sqrt{\frac{\hbar c}{G}} \sim 22 \mu \g
\eeq
In other words, microscopic and macroscopic values of mass $m$ may be 
delineated by comparing the associated Compton length $\lC$ to the Planck 
length $\lP$ 
\beq
m \lessgtr \mP &\Leftrightarrow & \lP \lessgtr \lC = \frac{\hbar }{mc} 
\eeq
It is then tempting to consider that this coincidence is not just 
accidental but that it might be a consequence of the existence of fundamental 
gravitational fluctuations. The idea was already present in the Feynman lectures 
on gravitation \cite{Feynman} and it was developed and 
popularized by a number of authors, for example 
\cite{Karolyhazy66,Diosi89,Penrose96}. 
The results obtained for gravitationally induced decoherence allow one
to test quantitatively this idea. 

To this aim, we rewrite the phase variance which determines decoherence
in all the systems studied in this paper as
\beq
\left\langle e ^{i \delta \Phi} \right\rangle 
= \exp \left( -\frac {\Delta \Phi^2}{2} \right) \ && \  
\frac{\Delta \Phi^2}{2} \simeq \mu^2 \ S_{h} \ \tau 
\eeq
Introducing the squared Planck time 
\beq
&&\tP ^2 = \frac{\hbar G}{c^5} = \left( \frac{\hbar}{\mP c^2} \right)^2
\eeq
we express the gravitational spectral density $S_{h}$ as
\beq
&&S_{h} \simeq \Theta _\gw  \ \tP ^2
\eeq
where $\Theta _\gw $ is a frequency measuring the temperature of the
background 
\beq
&&\Theta _\gw  \simeq \frac{\kB T }{\hbar} \simeq 10^{52}\s^{-1}
\eeq
Note that terms of order of unity are disregarded in these scaling arguments.

Collecting these relations, we obtain 
\beq
&&\frac{\Delta \Phi ^2}{2} \simeq \left(\frac{mv^2\sin\alpha}{\mP c^2} \right)^2 \ 
\Theta _\gw \tau 
\label{DeltaPhi2}
\eeq
The ratio $\frac{m^2}{\mP^2}$ which appears in this expression suggests that 
the Planck mass effectively plays a role in the definition of
the borderland between microscopic and macroscopic masses. 
However the presence of the other terms in the formula 
implies that the scaling argument on masses is not sufficient for obtaining
quantitative estimations. The phase variance also depends on the ratio of 
the probe velocity over the velocity of light, on the equivalent aperture angle
$\alpha$ and on the frequency $\Theta _\gw$ which measures gravitational noise level 
at the frequency of interest for the motion under study. This last quantity
has an enormous value, so that the transition between quantum and classical
regimes could in principle be observed for masses smaller than Planck mass.
Note that the parameter to be compared with Planck energy $\mP c^2$ is the
kinetic energy $m v^2$ of the probe rather than its mass energy $m c^2$.

Formula (\ref{DeltaPhi2}) can be used to answer the question whether or not it is 
possible to find systems on which the quantum/classical transition induced by 
intrinsic gravitational fluctuations can be observed experimentally. 
In order to approach the transition region $\Delta \Phi ^2 \sim 1$, one has
to consider heavy and fast enough particles in a matter-wave interferometer. 
Interference patterns have already been observed on fullerene molecules
\cite{Nairz01}. The kinetic energy of these molecules, 
the area and aperture angle of the interferometer are such that the 
gravitational decoherence has a negligible effect in these experiments,
as in HYPER. Increasing these numbers so that the transition
could be approached appears to be a formidable experimental challenge
with current technology.
An alternative approach is to look at interferometers using quantum condensates 
(see for example \cite{Anandan81,Chiao82} for suggestions along these lines) 
but this requires new experimental developments 
(see for example \cite{vanderWal00,Kasevich01,Varoquaux01,Vion02})
as well as new theoretical ideas.

\section*{Acknowledgments}

P.A.M.N. wishes to thank CAPES, CNPq,
FAPERJ, PRONEX, COFECUB, ENS and MENRT for their financial support which
made possible his stays in Paris during which this work was performed.

\end{document}